\documentclass[useAMS,usenatbib]{mn2e}

\usepackage{graphicx}
\usepackage{amsmath}
\usepackage{amssymb}
\usepackage{color}
\usepackage{subfig}
\usepackage[breaklinks,colorlinks,citecolor=blue,linkcolor=red]{hyperref} 
\usepackage[all]{hypcap}

\newcommand\msun{\, \rm M_\odot}

\newcommand\kms{\, \rm km\,s^{-1}}
\newcommand\pc{{\, \rm pc}}

\newcommand\yr{{\, \rm yr}}

\newcommand\eout{{e_{\rm out}}}

\newcommand\aout{{a_{\rm out}}}
\newcommand\ain{{a_{\rm in}}}
\newcommand\mimbh{{M_{\rm IMBH}}}

\title[Eccentric BBH mergers in GCs hosting IMBH]{Eccentric binary black hole mergers in globular clusters hosting intermediate-mass black holes}
\author[G. Fragione \& O. Bromberg]{  \parbox{\textwidth}{Giacomo Fragione$^{1}$\thanks{E-mail: giacomo.fragione@mail.huji.ac.il}, Omer Bromberg$^{2}$ \vspace*{0.3cm}}\\
$^1$Racah Institute for Physics, The Hebrew University, Jerusalem 91904, Israel\\
$^2$The Raymond and Beverly Sackler School of Physics and Astronomy, Tel Aviv University, Tel Aviv 69978, Israel}

\hypersetup{draft}

\begin{document}

\maketitle

\begin{abstract}
Globular clusters (GCs) may harbour intermediate-mass black holes (IMBHs) at their centres. In these dynamically active environments stellar-mass black holes (SBHs) sink to the center soon after formation, due to dynamical friction and start interacting among themselves and with the central IMBH. Likely, some of the SBHs will form bound systems with the IMBH. A fraction of those will be triple systems composed of binary SBHs and the IMBH acting as a third distant perturber. If the SBH binary orbit is sufficiently inclined it can develop Lidov-Kozai (LK) oscillations, which can drive the system to high eccentricities and eventually to a merger due to gravitational wave (GW) emission on short timescales. In this work, we focus on the dynamics of the IMBH-SBH-SBH triples and illustrate that these systems can be possible sources of GWs. A distinctive signature of this scenario is that a considerable fraction of these mergers are highly eccentric when entering the LIGO band ($10$ Hz). Assuming that $\sim 20\%$ of GCs host IMBHs and a GC density in the range $n_{_{\rm GC}}=0.32$-$2.31\,\mathrm{Mpc}^{-3}$, we have estimated a rate $\Gamma=0.06$-$0.46\,\mathrm{Gpc}^{-3}\,\mathrm{yr}^{-1}$ of these events. This suggests that dynamically-driven binary SBH mergers in this scenario could contribute to the merger events observed by LIGO/VIRGO. Full $N$-body simulations of GCs harbouring IMBHs are highly desirable to give a more precise constrain on this scenario.
\end{abstract}


\section{Introduction}

Black holes are divided into three categories according to their masses. (i) Stellar-mass black holes (SBHs) with typical masses of $10\msun\lesssim M\lesssim 100\msun$, are the remnants of  massive stars. To present day, $20$ SBHs in merging binaries have been observed by the LIGO-Virgo collaboration \citep{ligo2018}. (ii) Supermassive black holes (SMBHs) having masses $M\gtrsim 10^5 \msun$, reside in the centres of galaxies and shape the surrounding gas and the stellar distributions \citep{kor2013,alex17}, as observed in our Galaxy \citep{ghez2008,grav2018}. (iii) Intermediate-mass black holes (IMBHs) with masses $100\msun\lesssim M\lesssim 10^5 \msun$, which have been postulated to form through several mechanisms \citep[e.g.][]{por02,mckernan12,gie15}. Though we still lack a concrete proof of their existence, a recent observation of a tidal disruption event in an off-centre stellar cluster, consistent with an IMBH of $\sim 5\times 10^4 \msun$,  provides a strong supporting evidence of their existence \citep{lin18}. 

A natural place for IMBHs to reside is at the core of globular clusters (GCs), if the $M$--$\sigma$ relation observed in SMBHs 
is valid also in the IMBHs mass range \citep{por04,mer13,fgk18}. Galactic nuclei may host IMBHs as well, possibly delivered by inspiraling stellar clusters \citep*{mas14,agu18,flgk18} or formed \textit{in-situ} \citep{mckernan12}. They may merge with SBHs via gravitational waves (GWs) emission, as intermediate-mass ratio inspirals (IMRIs) \citep{frlei18}.

Several attempts have been made in modeling the dynamics of GCs hosting IMBHs through direct $N$-body simulations \citep{baum05,lut13,leigh14,bau17}. These simulations rarely include self-consistently the angular momentum and gravitational energy losses via GWs emission \citep{kon13,has16,mac2016}. Moreover, a large initial population of binaries have been proven hard to simulate with $N$-body models, in particular when dealing with massive clusters and massive IMBHs \citep{tre07,subr2019}.

Previous studies have focused on understanding the properties and the rates of IMRIs in GCs, as one of the most promising sources of GWs in the LISA band frequencies \citep[e.g.][]{mang2009,mill2009,ama2018}. \citet{mand2008} discussed the possibility that IMBH-SBH binaries may merge as a consequence of cumulative interactions with other stars in the cluster, or due to Lidov-Kozai (LK) oscillations whenever a third body is bounded to the IMBH-SBH binary. Recently, \citet*{fgk18} and \citet*{flgk18} used a semi-analytic approach to calculate cosmological rates of IMRIs in an evolving population of GCs. 

GCs are favorable locations for merger of binary SBHs, which should be quite abundant in such  dense stellar environments. If a SBH binary (SBHB) forms a bound system with the IMBH, its eccentricity and inclination can oscillate due to the LK mechanism whenever the initial SBHB orbit is sufficiently inclined \citep{lid62,koz62}, similarly to what happens in galactic nuclei \citep{antoper12,fragrish2018,grish2018,hoan18}. If the number of binary SBH interacting with the IMBH, is large enough, the IMBH-SBHB scenario may contribute to the overall SBH merger rate predicted by other channels, which is estimated to be in the range $0.1$--$100$ Gpc$^{-3}$ yr$^{-1}$ \citep[][]{Belczynski2006,ant17,askar17,baner18,gimap18,rod18,sam18,frak18,michaely2019}.

\begin{figure} 
\centering
\includegraphics[scale=0.425]{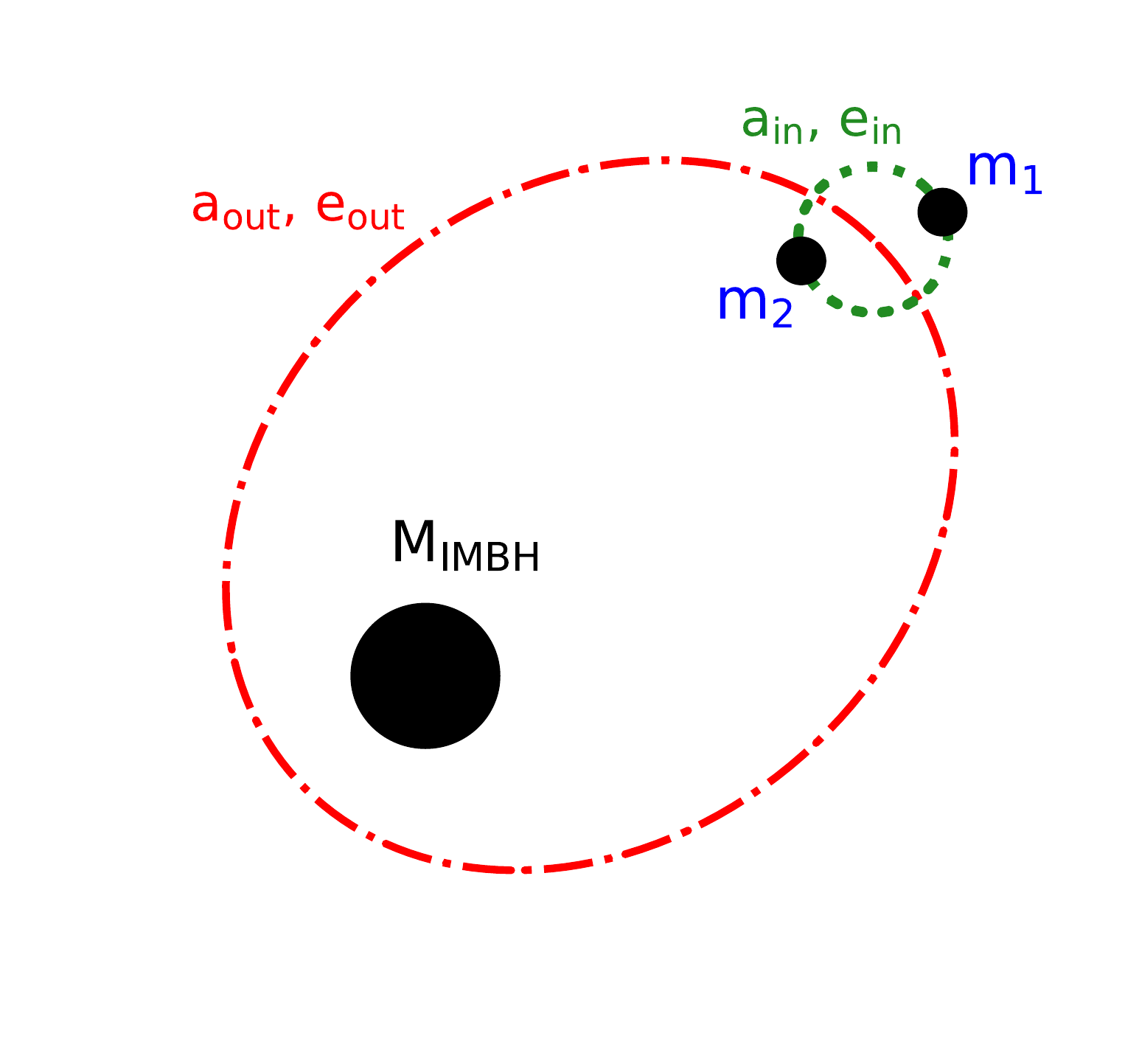}
\caption{The three-body system studied in the present work. We denote the mass of the IMBH as $M_\mathrm{IMBH}$ and the mass of the components of the SBHB binary as $m_1$ and $m_2$. The semimajor axis and eccentricity of the SBHB inner orbit are $a_{\rm in}$ and $e_{\rm in}$, respectively, while for the outer orbit $a_{\rm out}$ and $e_{\rm out}$.}
\label{fig:threebody}
\end{figure}

In this paper, we discuss the dynamics of triple systems made up of the central IMBH and a SBHB (see Figure~\ref{fig:threebody}). We show that the SBHB can undergo repeated LK oscillations during which its eccentricity becomes as large as unity. The SBHB will merge soon after formation, due to efficient dissipation via GW emission at pericenter. We sample different distribution masses of the SBH population, calculate the merger fraction and deduce merger rates for different IMBH masses.

The paper is organized as follows. In Section~\ref{sect:dynam}, we discuss the properties and dynamics of SBHs in the core of GCs harbouring IMBHs. In Section~\ref{sect:klmergers}, we present our numerical methods to determine the rate of IMBH-induced SBHB mergers and discuss the results. Finally, in Section~\ref{sect:conc}, we discuss the implications of our findings, compare them to a similar scenario in galactic nuclei, and draw our conclusions.

\section{Dynamics of stellar black holes near an intermediate-mass black hole}
\label{sect:dynam}

Soon after the cluster is created, massive stars collapse and form SBHs with masses that depend on the progenitors masses and metallicities \citep{bel2016,gimap18}. Assuming a canonical \citet{kro01} initial mass function, the number of SBHs is roughly proportional to the initial cluster mass with a coefficient 
\begin{equation}
N_{\mathrm{SBH}}\approx 3\times 10^{-3} \left(\frac{M_{\mathrm{GC}}}{\mathrm{M}_{\odot}}\right)\ .
\label{eqn:nsbh}
\end{equation}
The SBHs segregate towards the GC center on a dynamical friction timescale of \citep{binntrem87}
\begin{equation}
t_{\mathrm{seg}}\approx \frac{\bar{m}}{M_{\mathrm{SBH}}} t_{\mathrm{rh}}\ ,
\label{eqn:tdf}
\end{equation}
where $\bar{m}$ and $M_{\mathrm{SBH}}$ are the average stellar mass and SBH mass respectively and $t_{\mathrm{rh}}$ is the half-mass relaxation time:
\begin{equation}
t_{\mathrm{rh}}=54 \left(\frac{N}{10^5}\right)^{1/2} \left(\frac{r_h}{1\pc}\right)^{3/2} \left(\frac{1\msun}{\bar{m}}\right)^{1/2}\ \mathrm{Myr} \ .
\end{equation}
In the above equation $r_h$ is the cluster half-mass radius and $N$ is the number of objects in the cluster. For a typical cluster  $t_{\mathrm{rh}}\sim 10^8\yr$ and the SBHs segregate to the centre on a time scale of $t_{\mathrm{seg}}\sim 10^{6}$--$10^{7}\yr$, for $M_{\mathrm{SBH}}=10\msun$-$100\msun$, respectively. 

Traditionally, the segregated SBHs were thought to be decoupled from the rest of the cluster. They were believed to undergo strong gravitational interactions with each other which lead to their complete ejection out of the cluster \citep{spi69}. In practice the SBH sub-system is only partially decoupled from the cluster. Recent Monte-Carlo simulations by \citet{mor13} have found no evidence for the classical Spitzer instability, with only the innermost few tens of SBHs segregating significantly, while the majority remaining well mixed with the rest of the cluster. As a consequence, fewer than $\sim 50\%$ of the SBHs ($\sim 20$-$25\%$ of which are in binaries) will be ejected due to dynamical interactions at the cluster core.

In stellar clusters hosting IMBHs, typically one of the segregated SBHs forms a bound pair with the central IMBH. Using $N$-body simulations, \citet{leigh14} showed that the formation time for such a pair $\lesssim 100$ Myr, for clusters of masses  $2.0\times 10^4\msun$--$8.0\times 10^4\msun$. Generally, the IMBH can acquire a companion by capturing a SBH via the two-body capture process by gravitational radiation on a timescale \citep{miller2002}
\begin{eqnarray}
\tau_\mathrm{2,cap}&=&1.6\left(\frac{10^5\ \mathrm{pc}^{-3}}{n}\right)\left(\frac{30\msun}{M_{\mathrm{SBH}}}\right)^{11/7}\times\nonumber\\
&\times&\left(\frac{1000\msun}{M_{\mathrm{IMBH}}}\right)^{12/7}\left(\frac{\sigma_\mathrm{d}}{10\kms}\right)^{11/7}\ \mathrm{Myr}\ ,
\end{eqnarray}
or from tidally disrupting a SBHB, and capturing one of the SBH in the binary, via a three-body encounter on a timescale \citep{miller2002}
\begin{eqnarray}
\tau_\mathrm{3,enc}&=& 0.6\left(\frac{0.1}{\eta}\right)\left(\frac{1000\msun}{M_{\mathrm{IMBH}}}\right)\left(\frac{10^5\ \mathrm{pc}^{-3}}{n}\right)\times\nonumber\\
&\times& \left(\frac{\sigma_\mathrm{d}}{10\kms}\right) \left(\frac{10\ \mathrm{AU}}{a_{\rm SBHB}}\right)\ \mathrm{Myr}\ .
\end{eqnarray}
In the above equations, $n$ is the stellar number density, $\sigma_\mathrm{d}$ their velocity dispersion, $M_{\mathrm{IMBH}}$ and $M_{\mathrm{SBH}}$ are the masses of the IMBH and SBH, respectively, $\eta$ is the fraction of SBHs in binaries, and $a_{\rm SBHB}$ is the semi-major axis of the SBHB that interacts with the IMBH. Shortly after formation, the IMBH-SBH binary usually has a very high eccentricity, which leads to a decrease of the semi-major axis due to dynamical friction. Later, the binary interacts with ambient stars and compact objects via scattering slingshots at the typical hardening radius \citep{mer13}
\begin{equation}
a_{\mathrm{h}}=\frac{M_{\mathrm{SBH}}}{M_{\mathrm{IMBH}}+M_{\mathrm{SBH}}} \frac{r_{\mathrm{inf}}}{4}\ ,
\end{equation} 
where $r_{\mathrm{inf}}=GM_{\mathrm{IMBH}}/\sigma_d^2$ is the influence radius of the IMBH. Typically, $a_{\mathrm{h}}$ ranges from a few AU in the most massive clusters to a few hundreds AU in the lightest ones, which have smaller velocity dispersions and less massive IMBHs. 

As the IMBH-SBH binary hardens and the semi-major axis decreases, it becomes comparable to the typical radius over which gravitational radiation becomes important: \citep{has16}
\begin{eqnarray}
a_{_\mathrm{GW}}&\approx &\frac{0.12\ \mathrm{AU}}{(1-e^2)^{7/10}}\left(\frac{M_{\mathrm{IMBH}}}{1000\ \mathrm{M}_{\odot}}\right)^{1/5}\left(\frac{M_{\mathrm{SBH}}}{30\ \mathrm{M}_{\odot}}\right)^{1/5}\times\nonumber\\
&\times & \left(\frac{\sigma_d}{10\ \mathrm{km s}^{-1}}\right)^{1/5}\left(\frac{10^5\ \mathrm{pc}^{-3}}{n}\right)^{1/5}\ .
\label{eqn:agw}
\end{eqnarray}
The typical time to the next interaction falls below the GW timescale. Thus the IMBH-SBH merge within a \citet{pet64} timescale
\begin{equation}
T_{\mathrm{GW}}=\frac{3}{85}\frac{A^4 c^5}{G^3 M_{\mathrm{IMBH}} M_{\mathrm{SBH}} M}(1-E^2)^{7/2}\ ,
\label{eqn:peters}
\end{equation}
where $A$ and $E$ are the binary semi-major axis and eccentricity, respectively, and $M=M_{\mathrm{IMBH}}+M_{\mathrm{SBH}}$. Using $N$-body simulations, \citet{leigh14} showed that $T_{\mathrm{GW}}$ is typically at the range $\sim 10^5-10^9$ yr. If this happens and the merger product is not ejected due to GW recoil kick \citep*{holl2008,kon13,fgk18,flgk18}, it will capture another SBH, commonly less massive than the previous one, and the new-born binary will undergo the dynamical phases previously described.

SBHBs are frequently formed in the core of the cluster due to the high stellar density there \citep[see e.g.][]{rod16}. Since they are more massive than single SBHs ($m_{\rm SBHB}=m_1+m_2$), they sink more rapidly (see Eq.~\ref{eqn:tdf}) and eventually form bound triple systems with the central IMBH. This process can happen on a typical timescale $\sim \tau_\mathrm{2,cap}/\eta$. If the SBHB orbits the IMBH in a plane with inclination $i_0\sim 40^\circ$-$140^\circ$ with respect to the SBHB inner orbital plane, Lidov-Kozai (LK) cycles can influence the SBHB dynamics \citep{lid62,koz62}. In this scenario, the SBHB eccentricity oscillates on a typical timescale \citep{antognini15,nao16}
\begin{equation}
T_{\rm LK}=\frac{8}{15\pi}\frac{m_{\rm tot}}{M_{\rm IMBH}}\frac{P_{\rm IMBH}^2}{P_{\rm SBHB}}\left(1-e_{\rm out}^2\right)^{3/2}\ ,
\end{equation}
where $m_{\rm tot}=m_{\rm SBHB}+M_{\rm IMBH}$, and $P_{{\rm SBHB}}$ and $P_{{\rm IMBH}}$ are the orbital periods of the inner and outer orbits, respectively. Assuming $\eout=0.5$,
\begin{eqnarray}
T_{\rm LK}&=& 0.02\ \mathrm{Myr}
\left(\frac{\aout}{1000\ \mathrm{AU}}\right)^3\left(\frac{10\ \mathrm{AU}}{\ain}\right)^{3/2}\times\nonumber\\
&\times& \left(\frac{m_{\rm tot}}{1020\msun}\right)\left(\frac{1000\msun}{\mimbh}\right)^2\left(\frac{M_{\rm SBHB}}{30\msun}\right)^{1/2}\ .
\end{eqnarray}
At the quadrupole order of approximation (inner test particle and outer circular orbit), the maximal eccentricity is simply a function of the initial mutual inclination for an initial circular inner orbit
\begin{equation}
e_{\rm in}^{\rm max}=\sqrt{1-\frac{5}{3}\cos i_0^2}\ ,
\label{eqn:emax}
\end{equation}
which approaches unity as $i_0$ approaches $\sim 90^\circ$. In the case the outer orbit is eccentric and $m_1\neq m_2$, the inner eccentricity can reach almost unity even if the initial inclination is outside of the LK-angle range \citep[octupole order of approximation;][]{naoz13a}. This happens over the octupole timescale
\begin{equation}
T_{\rm oct}=\frac{1}{\epsilon}T_{\rm LK}\ ,
\label{eqn:tlkoct}
\end{equation}
where the octupole parameter is defined as 
\begin{equation}\label{oc1}
\epsilon={m_1-m_2\over m_1+m_2}\frac{\ain}{\aout}\frac{\eout}{1-e_{\rm out}^2}\ .
\end{equation}
Assuming $m_1=20\msun$ and $m_2=10\msun$, $\ain=10$ AU, $\aout=1000$ AU, and $\eout=0.5$, $T_{\rm oct}\sim 10$ Myr.

In some configurations, LK cycles can be suppressed by relativistic precession \citep{naoz2013,nao16}, which operates on a timescale
\begin{equation}
T_{\rm GR}=\frac{a_{\rm in}^{5/2}c^2(1-e_{\rm in}^2)}{3G^{3/2}(m_{\rm SBHB})^{3/2}}\ .
\end{equation}
Assuming inner circular orbit,
\begin{equation}
T_{\rm GR}=10\ \mathrm{Myr}\left(\frac{\ain}{10\ \mathrm{AU}}\right)^{5/2}\left(\frac{30\msun}{m_{\rm SBHB}}\right)^{3/2}
\end{equation}
In the region of the parameter space where $T_{\rm LK}>T_{\rm GR}$ and $T_{\rm oct}>T_{\rm GR}$, the LK oscillations, at quadrupole and octupole order, respectively, of the SBHB orbital elements are quenched by relativistic effects. However, if $T_{\rm LK}\sim T_{\rm GR}$ or $T_{\rm oct}\sim T_{\rm GR}$, a resonant-like behaviour may occur and LK cycles can be triggered, possibly exciting the inner orbit eccentricity to larger values \citep{naoz2013}.

Finally, binaries may evaporate due to dynamical interactions with stars (of average mass $\bar{m}$) in the core of the cluster. This happens on an evaporation timescale \citep{binntrem87}
\begin{eqnarray}
T_{\rm EV}&=& 30\ \mathrm{Myr} \left(\frac{m_{\rm SBHB}}{30\msun}\right)\left(\frac{\sigma}{10\ \mathrm{km s}^{-1}}\right)\times\nonumber\\
&\times&\left(\frac{1\ \msun}{\bar{m}}\right)\left(\frac{10\ \mathrm{AU}}{a_{\rm in}}\right)\left(\frac{10^5\ \mathrm{pc}^{-3}}{n}\right)\left(\frac{15}{\ln \Lambda}\right)\ .
\label{eqn:binevap}
\end{eqnarray}

\begin{table}
\caption{Simulated Models. Columns (from the left): IMBH mass ($M_\mathrm{SMBH}$), slope of the BH mass function ($\beta$), slope of the outer semi-major axis distribution ($\alpha$), merger fraction ($f_{\rm merge}$).}
\centering
\begin{tabular}{cccc}
\hline
$M_\mathrm{IMBH}$ (M$_\odot$) & $\beta$ & $\alpha$ & $f_{\rm merge}$ (\%)\\
\hline\hline
$1\times 10^3$ & $1$ & $2$     & $1.25$\\
$1\times 10^4$ & $1$ & $2$     & $1.55$\\
$1\times 10^5$ & $1$ & $2$     & $2.13$\\ 
$1\times 10^3$ & $2$ & $2$     & $1.23$\\
$1\times 10^3$ & $3$ & $2$     & $1.47$\\
$1\times 10^3$ & $4$ & $2$     & $1.14$\\
$1\times 10^3$ & $1$ & $1.5$   & $1.94$\\
$1\times 10^3$ & $1$ & $3$     & $3.93$\\
\hline
\end{tabular}
\label{tab:models}
\end{table}

\begin{figure} 
\centering
\includegraphics[scale=0.55]{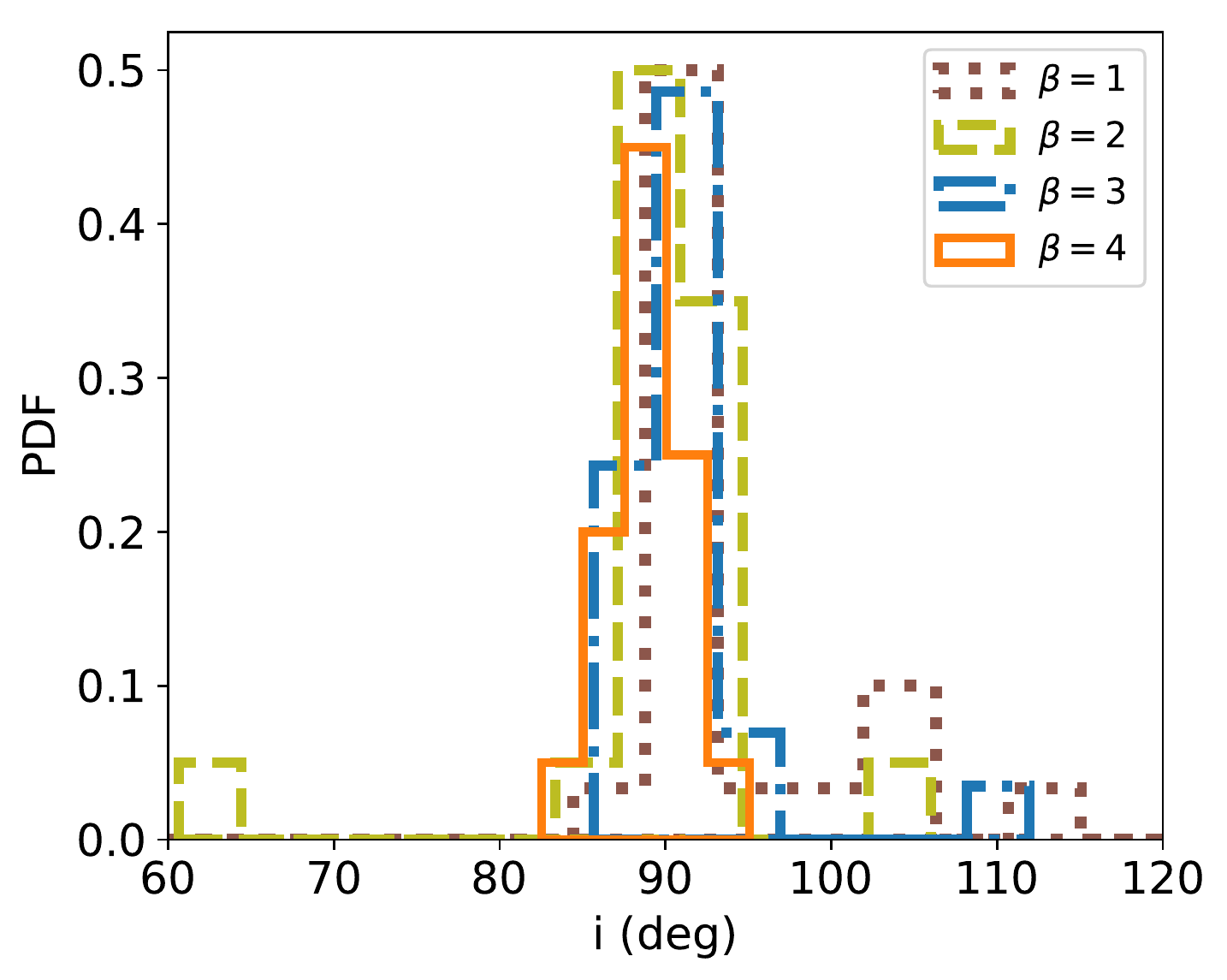}
\caption{PDF of the initial inclination of merging BH-BH binaries orbiting a $1 \times 10^3 \msun$ MBH, for different values of the slope of the BH mass function $\beta$ and $\alpha=2$.}
\label{fig:incl}
\end{figure}

Compared to the case the IMBH forms a binary system with a single SBH, the evolution of an IMBH-SBHB is dynamically rich \citep{chen18}. The large eccentricity values reached by the SBHB make its nominal Peters' merger time (Eq.~\ref{eqn:peters}) shorter, since it efficiently dissipates energy when $e \sim e_{\rm in}^{\rm max}$ \citep[e.g., see][]{antoper12}. Eventually, the SBHB is lead to a merger as a consequence of the GW radiation emitted at the pericenter. Even if the binary does not merge, it may appear in the the LISA frequency band and can be observed with a large enough signal-to-noise ratio \citep{hoa2019,rand2019}, thus possibly revealing the presence of the IMBH. 

The overall process that leads to the formation of a triple IMBH-SBHB depends mainly on the IMBH mass and on the core density of the host cluster, 
thus ultimately on the cluster mass if the IMBH mass correlates with it \citep{por02}. As discussed, the formation of an IMBH-SBHB triple can occur on a timescale $\sim \tau_\mathrm{2,cap}/\eta$, which is of the order of a few Myrs for typical GC parameters. Thus, the SBHB can be driven to merge by tidal interactions in the LK regime with the IMBH, before the binary orbit is perturbed by other SBHs and stars \citep{leighs2011}, and eventually evaporates, and before the IMBH interacts with different single or binary SBHs. \footnote{Some hard SBHB can form, and eventually merge, by interacting with other SBHs and SBHBs in the core of the GC, as in the usual scenario where there is no IMBH in its centre \citep{rod16}.}

In the next Section, we consider IMBH-SBHB triples formed through the processes described above. We integrate their equation of motions to quantify the fraction of mergers, and give an estimate of the possible SBH merger rate from this channel.

\section{Numerical simulations of stellar black hole mergers}
\label{sect:klmergers}

\subsection{Initial conditions}

To illustrate the efficiency of the IMBH-SBHB mechanism and to examine its dependence on the cluster properties, we perform high-precision $N$-body simulations of the dynamics of the triplet in GCs of with various central IMBHs and background stellar cusp objects. The properties of the stellar cusp will determine the typical time for the triplet to have strong interactions with a cusp object. We therefore set the total integration time to be the minimum between $\tau_\mathrm{2,cap}$ and $\tau_\mathrm{3,enc}$, an order of a $\sim$ few Myr. This also relies on the fact that on longer timescales ($\sim 10$ Myr) the LK oscillations could be suppressed by the GR precession, and on the fact that SBHBs may evaporate as a results of the dynamical interactions with other stars in the cluster core ($\sim 30$ Myr). The simulations are performed with the \textsc{archain} code \citep{mik06,mik08}. \textsc{archain} is a fully regularized code able to model the evolution of systems of arbitrary masses, radii and eccentricities with extreme accuracy, and includes Post-Newtonian (PN) corrections up to order PN2.5.

We consider three different masses for the IMBH: $\mimbh=10^3\msun$, $10^4\msun$, $10^5\msun$. Stars and compact objects tend to form a power-law density cusp ($n(r)\propto r^{-\alpha}$) around an IMBH similar to galactic nuclei, where lighter (heavier) objects develop shallower (steeper) cusps \citep{bahcall76}. Typically, stars tend to have $\alpha\sim 1.5$-$1.75$, while SBHs $\alpha\sim 2$-$3$ as a result of mass segregation \citep{alex17,baumg18}. Therefore, we assume that the background SBH number density follows a cusp with $\alpha=2$. 
To study how the cusp slope affects the results, we run two additional models with a shallower cusp ($\alpha=1.5$) and a steeper cusp ($\alpha=3$). We take the maximum outer semimajor axis to be $0.1\times(\mimbh/4\times 10^6\msun)^{2/9}$ pc \citep{hoan18}\footnote{The maximum outer semimajor axis is chosen such that the octupole LK timescale is equal to the timescale on which accumulated fly-bys from single stars tend to unbind the binary.}.

We choose a negative power-law distribution for the SBH mass
\begin{equation}
\frac{dN}{dM} \propto M^{-\beta}
\label{eqn:bhmassfunc}
\end{equation}
in the mass range $5\msun$-$100\msun$\footnote{Recent theoretical results on pulsational pair instability limit the maximum mass to $\sim 50\msun$ \citep{bel2016}.} \citep{hoan18} and study how the results depend on the slope $\beta$ by running models with $\beta=1$, $2$, $3$, $4$ \citep{olea16}.

We assume that the distribution of the semi-major axes of the SBHB is flat in log-space (\"{O}pik's law), while the inner and outer eccentricities are drawn from a thermal distribution \citep{jeans1919}. The initial inclination $i_0$ between the plane of the SBHB and the pair's orbital plane around the IMBH is sampled from an isotropic distribution (i.e. uniform in $\cos i$). The other relevant angles are drawn randomly.

\begin{figure} 
\centering
\includegraphics[scale=0.55]{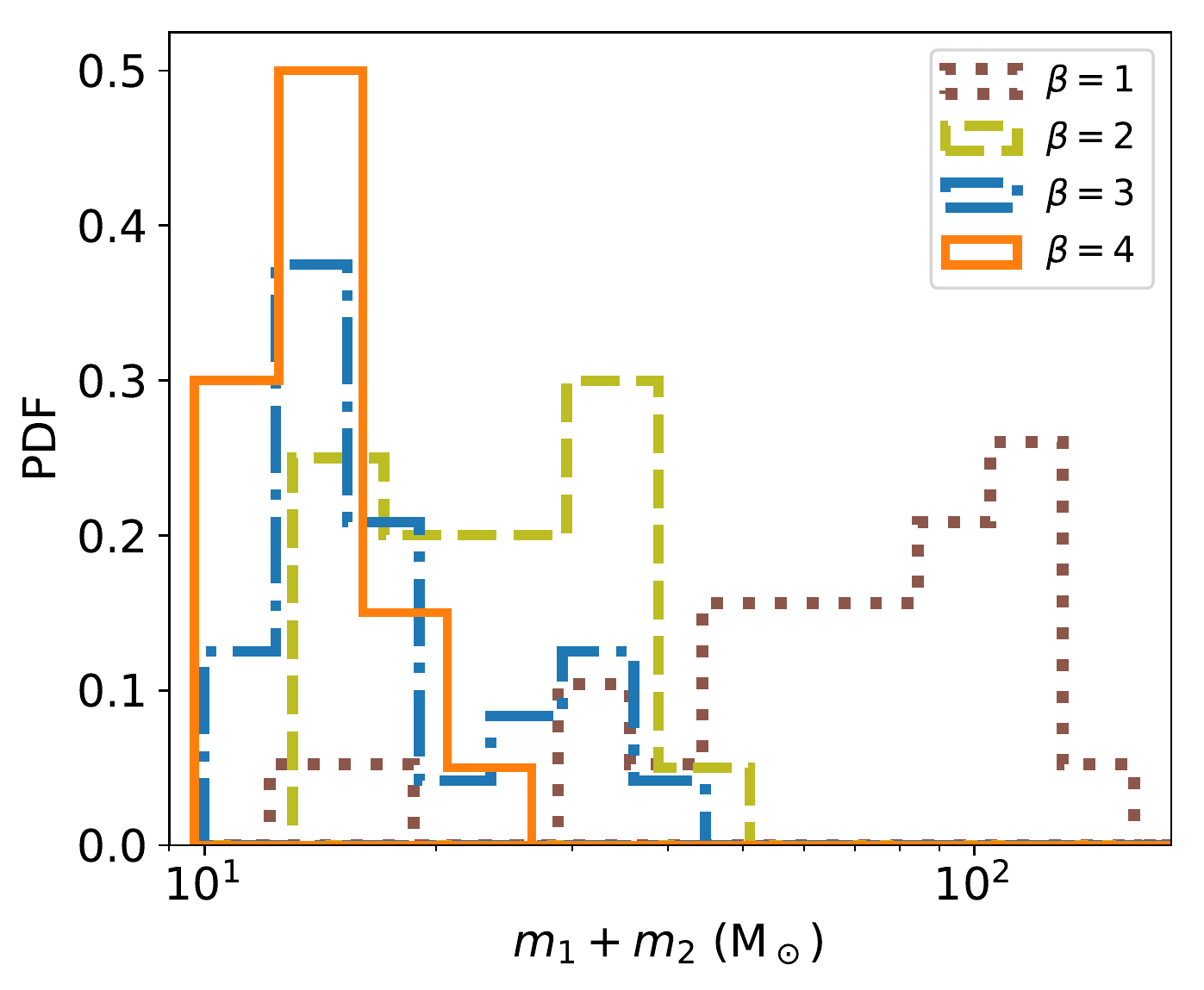}
\caption{Total mass PDF of merging BH-BH binaries orbiting a $1 \times 10^3 \msun$ MBH, for different values of the slope of the BH mass function $\beta$ and $\alpha=2$.}
\label{fig:mass}
\end{figure}

After we sample the initial orbital parameters from the relevant distributions, we check that the \citet{mar01} criterion\footnote{Other stability criteria \citep[e.g.][]{myll2018} could deem unstable a (slightly) different portion of the parameter space.}
\begin{equation}
\frac{R_{\rm p}}{a_{\rm in}}\geq 2.8 \left[\left(1+\frac{\mimbh}{m_1+m_2}\right)\frac{1+e_{\rm out}}{\sqrt{1-e_{\rm out}}} \right]^{2/5}\left(1.0-0.3\frac{i_0}{\pi}\right)\ ,
\label{eqn:stabts}
\end{equation}
is satisfied, where $R_{\rm p}$ is the pericenter of the outer orbit. If the system is stable, we start the integration, otherwise we sample again the relevant parameters of the system. Throughout the evolution of the triple system, the SBHB can undergo three fates: (i) it remains bound to the IMBH, possibly on an orbit perturbed with respect to the original one; (ii) it becomes unbound as a consequence of the tidal interactions with the IMBH; (iii) it merges as a consequence of GW emission, typically enhanced by the LK mechanism. The distinction among the first two cases is made by computing the mechanical energy of the SBHB with respect to the IMBH.

\begin{figure} 
\centering
\includegraphics[scale=0.55]{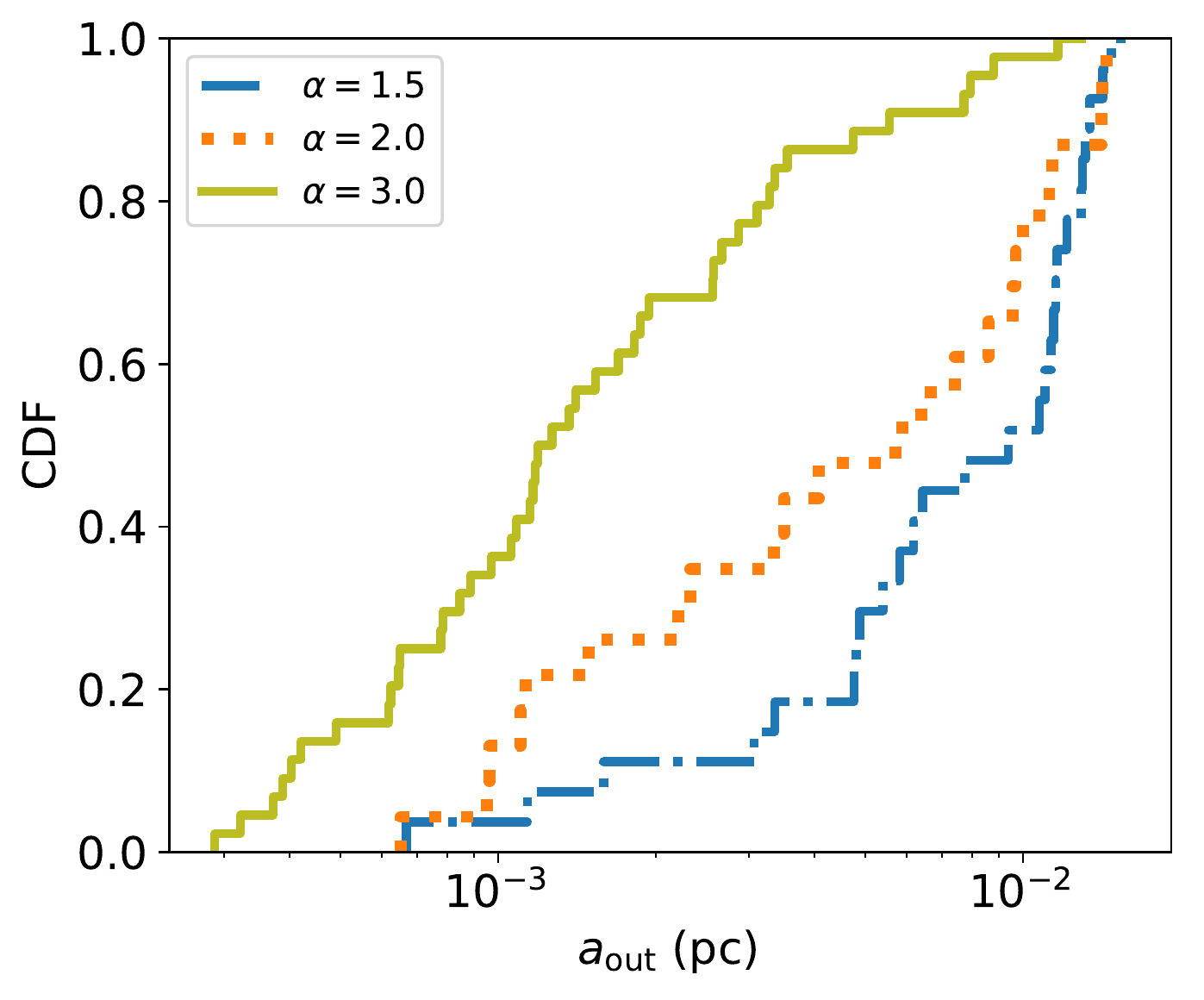}
\includegraphics[scale=0.55]{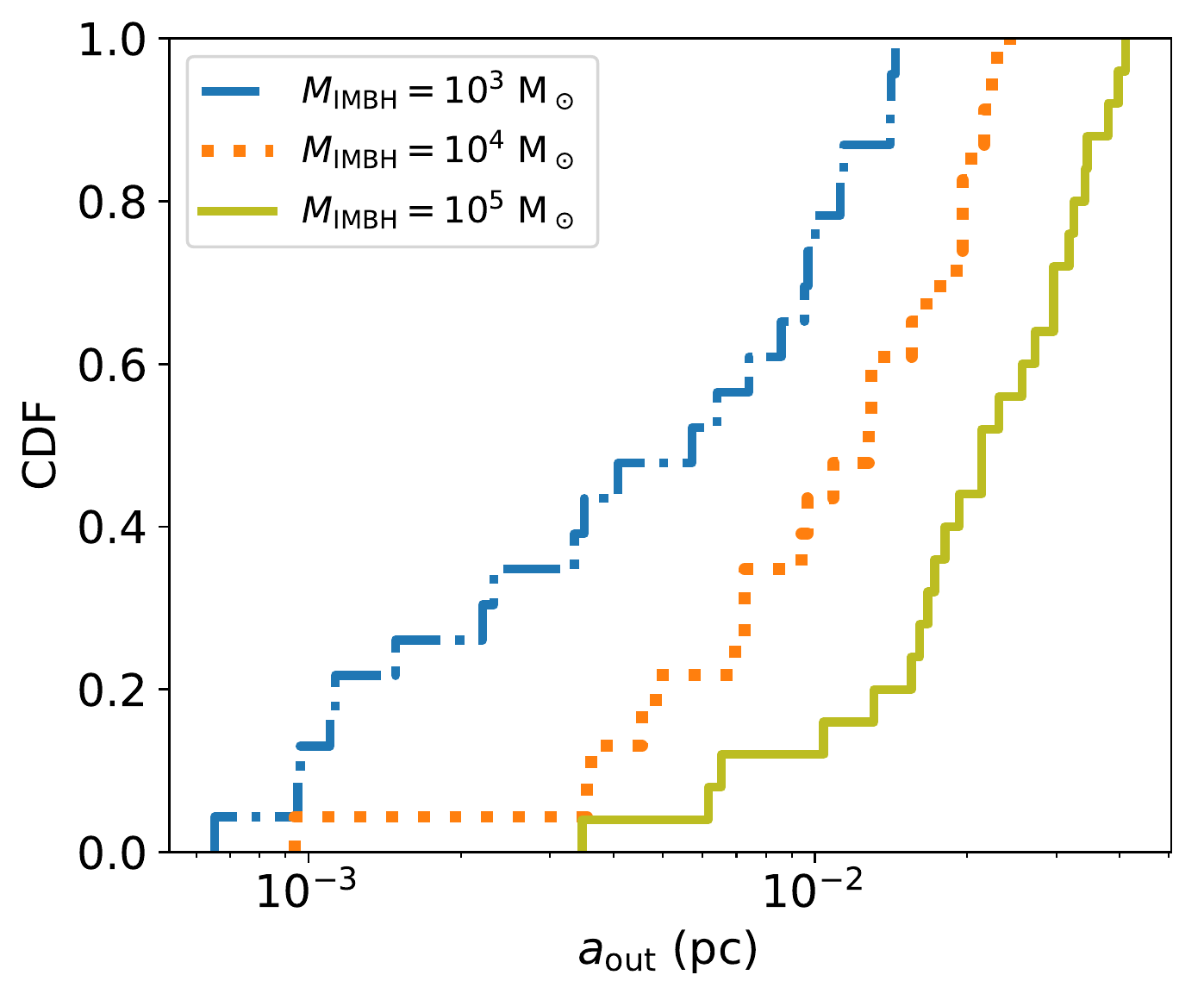}
\caption{Cumulative distribution functions of merging SBHBs outer orbits as a function of the slope of the SBHB cusp for $\mimbh=10^3\msun$ and $\beta=1$ (top panel) and as a function of the IMBH mass for $\beta=1$ and $\alpha=2$ (bottom panel).}
\label{fig:aout}
\end{figure}

\subsection{Results: inclination, total mass, semi-major axes and eccentricity distributions}

An SBHB is expected to be significantly perturbed by the tidal field of the IMBH whenever their mutual orbit is sufficiently inclined with respect to the orbital plane around the IMBH, $i_0\sim 40^\circ$-$140^\circ$  \citep{lid62,koz62}. According to Eq.~\ref{eqn:emax}, the SBHB eccentricity reaches almost unity when $i_0\sim 90^\circ$. Figure~\ref{fig:incl} shows the probability distribution function (PDF) of the initial inclination angle in those systems, which ended up in a merger. The distributions are shown for SBHs orbiting an $1 \times 10^3 \msun$ IMBH, having different values of $\beta$ and $\alpha=2$. Independently of the slope of the SBH mass function, the majority of the mergers take place when the initial inclination is $\sim 90^\circ$. In this case the LK effect is maximal, leading to eccentricity oscillates up to unity. The SBHB experiences rapid gravitational energy loss due efficient energy dissipation near the pericentre, which ends in a merger.

\begin{figure} 
\centering
\includegraphics[scale=0.55]{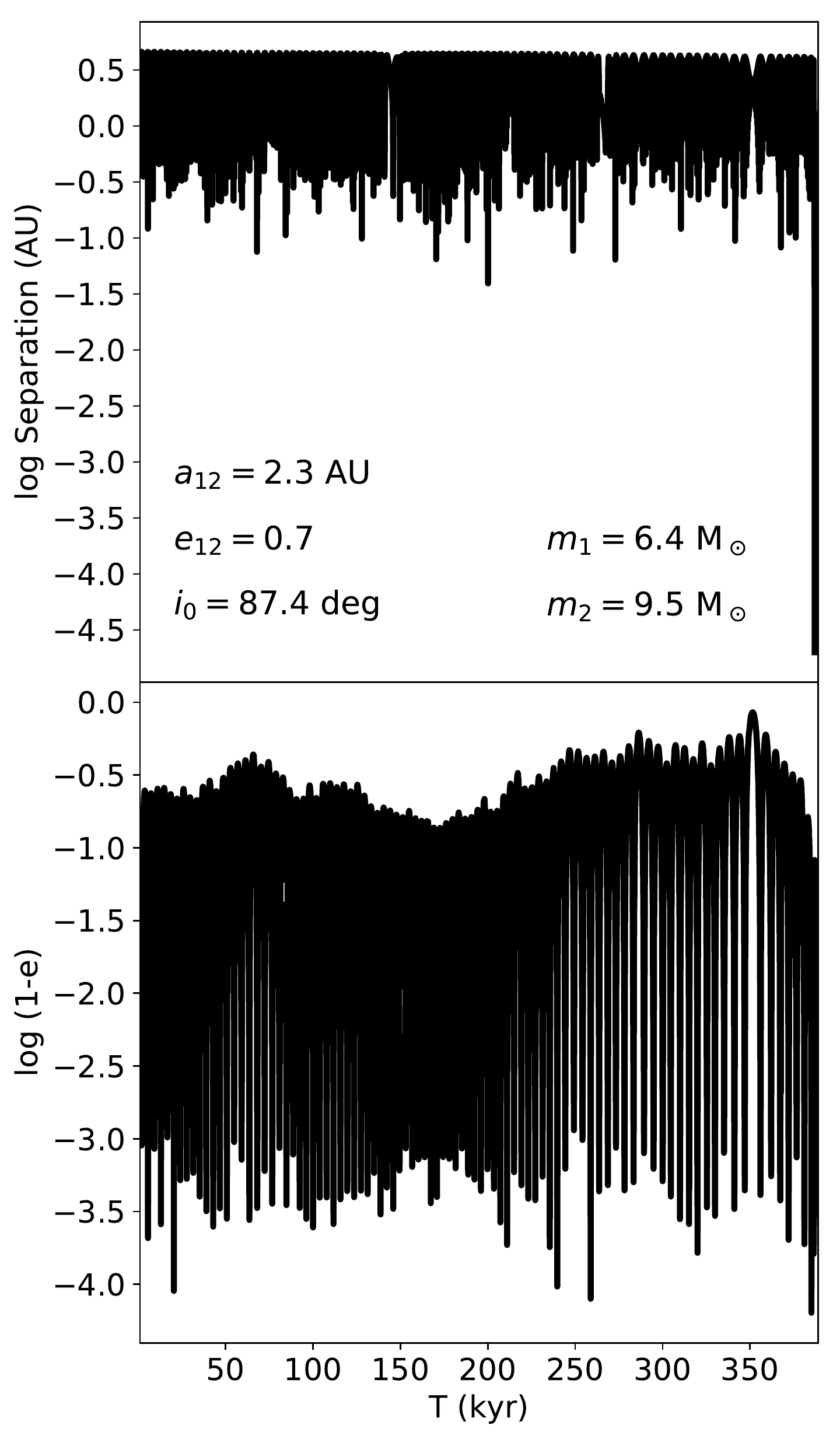}
\caption{Example of a three-body integration of an SBHB with masses $m_1=6.4\msun$ and $m_2=9.5\msun$ orbiting an IMBH of mass $\mimbh=10^3\msun$, which enter the LIGO frequency band with eccentricity $e_{10Hz}\sim 0.94$. The SBHB start with a semi-major axis $a_{12}=2.3$ AU, an eccentricity $e_{12}=0.7$ and an orbital inclination of $i_{12}=87.4^\circ$, and merge within $\sim 3.9\times 10^5$ yr.}
\label{fig:examecc}
\end{figure}

The slope of the SBH mass function, $\beta$, is largely unknown. To understand the effect it has on the masses of the SBHBs that undergo mergers we study four different slopes. Fig.~\ref{fig:mass} illustrates the mass distribution function of merging SBHs orbiting an $1 \times 10^3 \msun$ IMBH, for different values of $\beta$ and $\alpha=2$. Initial mass function with steeper slopes (larger $\beta$'s) lead to smaller masses of the merging SBHs, while shallower SBH mass functions (smaller $\beta$'s) favours more massive SBHs. We find that $90\%$ of the mergers have total masses smaller $\sim 100\msun$, $\sim 40\msun$, $\sim 30\msun$, $\sim 15\msun$ for $\beta=1$, $\beta=2$, $\beta=3$, $\beta=4$, respectively.

In Figure~\ref{fig:aout} (top panel), we report the cumulative distribution function (CDF) of merging SBHBs outer orbits as a function of $\alpha$ for $\mimbh=10^3\msun$ and $\beta=1$. We find that if the slope of the SBH cusp is shallower (smaller $\alpha$'s), SBHB merge at larger semi-major axes with respect to the orbit around the IMBH, on average. Additionally, the mass of the IMBH also affects the typical $\aout$ at which the SBHBs merge. Figure~\ref{fig:aout} (bottom panel) shows the CDF of merging SBHBs outer orbits as a function of the IMBH mass for $\beta=1$ and $\alpha=2$. SBHB marge typically closer to lighter IMBHs than heavier IMBHs. Lighter IMBHs have smaller influence spheres and SBHB have to be closer in order to avoid evaporation due to the interaction of surrounding stars and compact objects before merging due GW emission induced by LK oscillations.

\begin{figure} 
\centering
\includegraphics[scale=0.55]{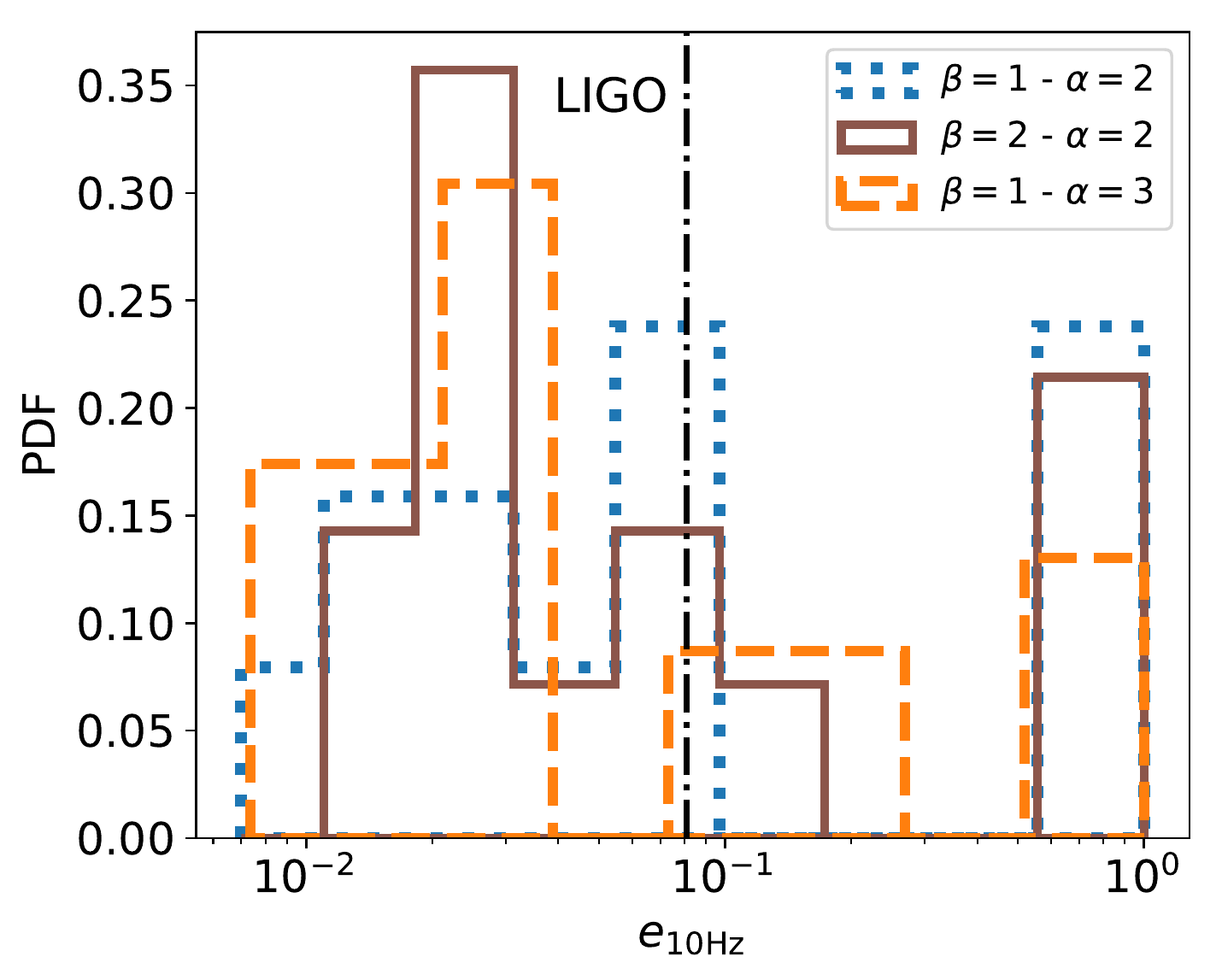}
\caption{Distribution of eccentricities at the moment the BH binaries enter the LIGO frequency band ($10$ Hz), for $\mimbh=10^3\msun$ and different values of $\beta$ and $\alpha$. We also show a vertical line at the level $e_{10Hz}=0.081$ where LIGO/VIRGO can start to detect sources \citep{gond2019}.}
\label{fig:ecc}
\end{figure}

SBHB in hierarchical configurations like IMBH-SBHB are expected to have large eccentricities in the LIGO frequency band ($10$ Hz), as a consequence of the perturbation by the third body and the LK cycles \citep[see e.g.][]{fragrish2018}. For the SBHBs that merge in our simulations, we compute a proxy to the GWs frequency, which we take to be the frequency corresponding to the harmonic that gives the maximal emission of GWs \citep{wen03}
\begin{equation} 
f_{\rm GW}=\frac{\sqrt{G(m_1+m_2)}}{\pi}\frac{(1+e_{\rm in})^{1.1954}}{[\ain(1-e_{\rm in}^2)]^{1.5}}\ .
\end{equation}
Figure~\ref{fig:examecc} shows an example of a three-body integration of an SBHB with masses $m_1=6.4\msun$ and $m_2=9.5\msun$ orbiting an IMBH of mass $\mimbh=10^3\msun$, which enter the LIGO frequency band with eccentricity $e_{10Hz}\sim 0.94$. The SBHB start with a semi-major axis $a_{12}=2.3$ AU, an eccentricity $e_{12}=0.7$ and an orbital inclination of $i_{12}=87.4^\circ$, and merge within $\sim 3.9\times 10^3$ yr. 

In Figure~\ref{fig:ecc}, we illustrate the distribution of eccentricities at the moment the SBHB enter the LIGO frequency band. We show results for mergers produced by SBHB with $\mimbh=10^3\msun$ and different values of $\beta$ and $\alpha$. SBHBs that merge through this channel have larger eccentricities than those formed through other channels, like mergers of isolated binaries or of SBHBs ejected from stellar clusters \citep{Belczynski2008,frak18,rod18}. Note that, mergers that follow from the GW capture scenario in clusters \citep{sams2014,sam18,ssdo2018,zevin18} and from hierarchical triples and quadruples \citep{ant17,fragk2019,fragl19} also present a similar shape and a similar peak at high eccentricities. We also show a vertical line at the level $e_{10Hz}=0.081$ where LIGO/VIRGO can start to detect sources \citep{gond2019}. Thus, highly-eccentric mergers might be an imprint of SBHBs that merge through this channel and can thus reveal the presence of an IMBH.

\subsection{Merger rates}

To accurately determine the global SBH merger rate from this channel, we would need to quantify the population of SBHBs that orbit an IMBH in GCs. A precise answer to this question would require running numerous $N$-body models of GCs harbouring IMBHs, as well as taking into account possible effects of the primordial binary fraction, degree of mass segregation, metallicity and so on, which is beyond the scope of the present paper. Nevertheless, we can use the simulations we have run in this work as a proxy for inferring an order-of-magnitude estimation to the merger rate of SBHBs interacting with IMBHs.

The merger rate of SBHB interacting with IMBH can be computed as
\begin{equation}
\Gamma=n_{_{\rm GC}} \zeta_{IMBH} \eta f_{\rm merge} \Gamma_{\rm sup}
\label{eqn:rates}
\end{equation}
In the previous equation, $n_{_{\rm GC}}$ is the GC density in the Universe, $\zeta_{IMBH}$ is the fraction of GCs that host an IMBH, $\eta$ is the SBHB fraction, $f_{\rm merge}$ is the fraction of SBHB that merge, and $\Gamma_{\rm sup}$ is the supply rate of SBHB. We assume a GC density in the range $n_{_{\rm GC}}=0.32$-$2.31\,\mathrm{Mpc}^{-3}$ \citep{por00,rod15} and $\zeta_{IMBH}\sim 0.2$ \citep{gie15}. In the last column of Table~\ref{tab:models}, we report the fraction of SBHB that merge in our simulations. Typically, it lies in the range $f_{\rm merge} \sim 1$--$4$\%, with higher fractions for more massive IMBHs and steeper cusp densities. Finally, $\Gamma_{\rm sup}$ is the supply rate of SBHB. As discussed, this happens as a consequence of either two-body capture process by gravitational radiation on a timescale $\tau_\mathrm{2,cap}/\eta$, typically of the order of $\sim$ Myr. Plugging numbers into Eq.~\ref{eqn:rates}
\begin{eqnarray}
\Gamma&=&0.2\ \mathrm{Gpc}^{-3}\ \mathrm{yr}^{-1} \left(\frac{n_{_{\rm GC}}}{1\ \mathrm{Mpc}^{3}}\right)\times\nonumber\\
&\times& \left(\frac{\zeta_{IMBH}}{0.2}\right)\left(\frac{\eta}{0.1}\right) \left(\frac{f_{\rm merge}}{0.01}\right) \left(\frac{\Gamma_{\rm sup}}{1\ \mathrm{Myr}}\right)
\end{eqnarray}

For reference, the merger rate in nuclear stellar clusters is at the range of $\sim 8.5$-$29.5\,\mathrm{Gpc}^{-3}\,\mathrm{yr}^{-1}$ \citep{ant16,petr17,fragrish2018,hamer18,hoan18,steph2019}, while the predicted merger rate from GCs is at the range of $\sim 1$-$20\,\mathrm{Gpc}^{-3}\,\mathrm{yr}^{-1}$ \citep{askar17,frak18,rod18}. Our results suggest that dynamically-driven SBH mergers in IMBH-SBHB systems may be important and could contribute to the merger events observed by LIGO/VIRGO, $9.7$--$101$ Gpc$^{-3}$ yr$^{-1}$ \citep[run O2;][]{ligo2018}. However, we caution that full $N$-body simulations of GCs harbouring IMBHs are highly desirable to provide a more precise constrain on the merger rate. Moreover, the estimated rates should be convolved with the GC history in a given galaxy to check how they evolve with redshift across cosmic time \citep*{fgk18,flgk18}. Finally, we note that hard SBHBs, that later merge, can form by interacting with other SBHs and SBHBs in the core of the GC, as in the usual scenario where there is no IMBH in its centre \citep{rod16}. Since we do not take this possibility into account, the above merger rate can be seen as a lower limit. Nevertheless, these SBHB will probably enter the LIGO band with a nearly circular orbit, unlike the case their merger is accelerated by the LK effect induced by the tidal field of the IMBH.

\section{Discussion and conclusions}
\label{sect:conc}

GCs may harbour IMBHs in their centres. In such a dynamical active environment, SBHs sink to the center soon after formation, due to dynamical friction and start interacting among themselves and with the central IMBH. Likely, some of these SBHs will form bound systems with the IMBH. If some of these SBHs are in binaries, the system they form with the IMBH is actually a triple, where the IMBH act as a third distant perturber. If the SBHB orbit is sufficiently inclined, it can develop LK oscillations which can drive the system to high eccentricities and merge due to GW emission on short timescales.

In this paper, we focus on the dynamics of IMBH-SBHB systems, that can  form in  cores of GCs. We illustrate how the LK mechanism operates in such a system. We consider different IMBH masses, adopting a mass spectrum for the BHs, and study different spatial distributions for the SBHB binaries. We show that the majority of systems merge when the SBHB orbital plane is initially inclined at $\sim 90^\circ$ with respect to the orbit of the SBHB around the IMBH, independent of the SBH mass function slope, $\beta$. However, $\beta$ controls the mass distribution of the merging SBHs, while the IMBH mass and the slope of the SBH cusp distribution ($\alpha$) control the distribution of the semi-major axis of merging SBHBs. A distinctive signature of this scenario is that a considerable fraction of mergers is highly eccentric when entering the LIGO band.

Although we still lacking confirmed evidence, GCs hosting IMBH should not be rare. Assuming that $\sim 20\%$ of clusters may host an IMBH, we have determined a typical merger rate of $\sim 0.2\ \mathrm{Gpc}^{-3}\ \mathrm{yr}^{-1}$. This merger rate is comparable to the SBHB merger rate in nuclear stellar clusters \citep{ant16,fragrish2018,hamer18}, but lower than the predicted merger rate from GCs \citep{askar17,frak18,rod18}. Nevertheless, our results suggest that dynamically-driven SBH mergers in IMBH-SBHB could not be rare and could be observed in the present and upcoming runs once hundreds of SBHB mergers will be detected. However, a more precise estimate would definitively require full $N$-body simulations of GCs harbouring IMBHs and should take into account the GC history in a given galaxy across cosmic time \citep*{fgk18,flgk18}.

As discussed, we have neglected the formation of hard SBHBs, that later merge, by interacting with other SBHs and SBHBs in the core of the GC, as in the usual scenario where there is no IMBH in its centre. \citep{rod16}, recently updated by \citet{rod18}, presented a large set of Monte Carlo models focusing on the formation and evolution of SBHBs though few-body interactions. Their models did not include any IMBH in their centre and did not take into account the possibility of forming IMBHs as a result of stellar mergers, as in \citet{gie15}. In the case no IMBH lurks in the center of a cluster, SBHs continuously interact and form binaries, which later can harden by means of the same interactions. In the case a GC hosts an IMBH in its centre, a similar picture could still be depicted for SBHs that reside outside of the sphere of influence of the IMBH. Within this sphere, the interactions are dominated by the IMBH gravitational field and the dynamics becomes more similar to the dynamics of galactic nuclei hosting SMBHs. The rate predicted in the dynamical-formation scenario in GCs (with no IMBH) is in the range of $1$-$20\,\mathrm{Gpc}^{-3}\,\mathrm{yr}^{-1}$ \citep{askar17,frak18,rod18}, thus larger than the IMBH-assisted merger. Nevertheless, the statistical contribution of different astrophysical channels can be hopefully disentangled using the spin, eccentricity, and redshift distributions \citep[e.g.][]{olea16}. The typical eccentricity distribution (at LIGO band) of SBHB mergers in GCs (with no IMBH) has a more complicated shape than the scenario proposed in this paper \citep{zevin18}. However, typically the SBHBs will enter the LIGO band with low eccentricities, while the majority of systems have a large eccentricity ($e_{\rm 10 Hz}\gtrsim 0.01$) if they merge as a result of the LK effect induced by the tidal field of the IMBH. Also, the effective spin measurable by LIGO could discriminate between the scenarios proposed in \citep{rod16} and in this paper. While in the former case even large effective spins could be observed \citep{rod18}, their distribution should be peaked towards zero in the case the mergers happens in a triple configuration \citep{anton2018}. Finally, for what concerns the redshift distribution, we expect a nearly similar distribution, since, both in the case the GC hosts and does not host an IMBH, the merger history is correlated to the GC history across cosmic time \citep{fgk18,frak18,rod18}.

A similar mechanism to the one studied here, has shown to occur at galaxies centers, where the SBHB interact with the SMBH \citep{antoper12,fragrish2018,hamer18,hoan18}. The estimated rates are roughly of the same order of magnitude, however their location in the host galaxy will be different. While the SBHB mergers driven by SMBHs will appear at the center of the galaxies, IMBH driven mergers will occur predominantly at the galactic bulges and halos. Thus, if given a good enough localization these events can be disentangled. Finally, we note that while neutron star-neutron star (NS-NS) and BH-NS mergers can happen through the same process near SMBHs, this should be uncommon in GCs harbouring IMBHs, where NSs are likely ejected due to birth kicks and interactions with SBHs \citep{fpb18}, thus preventing the formation of IMBH-NS-NS and IMBH-SBH-NS triplets.

\section*{Acknowledgements}

We thank the referee for a constructive report. GF is supported by the Foreign Postdoctoral Fellowship Program of the Israel Academy of Sciences and Humanities. GF also acknowledges support from an Arskin postdoctoral fellowship. Simulations were run on the \textit{Astric} cluster at the Hebrew University of Jerusalem.

\bibliographystyle{mn2e}
\bibliography{refs}

\end{document}